\documentclass[10pt,aps,prx,showpacs,superscriptaddress,preprint,citeautoscript]{revtex4-1}

\usepackage{graphicx,color}  
\usepackage{dcolumn}   
\usepackage{bm}        
\usepackage{amssymb,url}
\usepackage{amsmath}
\usepackage{hyperref}

\newcommand{\dbeq}{\begin{equation}}
\newcommand{\deeq}{\end{equation}}

\newcommand{\subeqs}[1]{\begin{subequations}\begin{align} #1 \end{align}\end{subequations}}
\def\dd{\operatorname{d}}

\def\eeq{\relax}
\def\beq#1#2\eeq{\begin{equation}\label{#1}#2\end{equation}}
\def\bal#1#2\eal{\begin{align}\label{#1}#2\end{align}}
\def\bse#1#2\ese{\begin{subequations}\label{#1}#2\end{subequations}}
\def\ba{\begin{aligned}}   \def\ea{\end{aligned}}
\def\tr{\operatorname{tr}} 
\def\dd{\operatorname{d}} 
\def\Im{\operatorname{Im}}
\def\Re{\operatorname{Re}} 
\def\C{C}
\def\p{\pi}

\def\rev#1{\textcolor{magenta}{#1}}

\begin{document}

\title{Loss Compensation in Time-Dependent Elastic Metamaterials}
\author{Daniel Torrent}
\email{dtorrent@uji.es}
\affiliation{Centre de Recherche Paul Pascal, UPR CNRS 8641, Universit\'e de Bordeaux, Pessac, (France)}
\affiliation{GROC, UJI, Institut de Noves Tecnologies de la Imatge (INIT), Universitat Jaume I, 12080, Castell\'o, (Spain)}
\author{William J.\ Parnell}
\affiliation{School of Mathematics, University of Manchester, Oxford Road, Manchester, M13 9PL, (United Kingdom)}
\author{Andrew N. Norris}
\affiliation{Mechanical and Aerospace Engineering, Rutgers University, Piscataway, NJ 08854-8058, (USA)}
\date{\today}

\begin{abstract}
Materials with properties that are modulated in time are known to display wave phenomena showing   energy increasing with time, with the rate mediated by the modulation.  Until now there has been no accounting for material dissipation, which clearly  counteracts energy growth. This  paper provides
an exact expression  for the amplitude of elastic or acoustic waves propagating in lossy materials with properties that are periodically modulated in time.  It is found that these materials can support a special propagation regime in which waves travel at constant amplitude, with temporal modulation compensating for the normal energy dissipation. We derive a general condition under which  amplification due to time-dependent properties  offsets  the material dissipation. This identity  relates band-gap properties associated with the temporal modulation and the average of the viscosity coefficient, thereby providing a simple recipe for the design of loss-compensated mechanical metamaterials.
\end{abstract}

\maketitle
\section{Introduction}
Phononic crystals and metamaterials, which consist of periodic arrangements of scatterers and resonators in a solid or fluid matrix, have revolutionized the realm of acoustic and elastic wave propagation\cite{craster2012acoustic,deymier2013acoustic,cummer2016controlling}. These structures have allowed the development of applications for the control and localization of mechanical energy that would be impossible to achieve with natural materials. Thus, gradient index lenses\cite{lin2009gradient,climente2010sound}, cloaking shells\cite{chen2007acoustic,cummer2007one,torrent2008acoustic} and hyperlenses\cite{li2009experimental}, among other interesting devices, have been designed and experimentally tested.

However, most of the extraordinary applications of metamaterials are hindered by the strong dissipation they exhibit, especially near the resonant regime where the concentration of the fields in the scatterers is higher\cite{pichard2016dynamic}. The performance of these structures could be considerably improved if combined with materials with gain. Additionally, other emerging applications related with PT-symmetric systems\cite{zhu2014p,christensen2016parity}, where gain and loss are combined, requires the realization of materials with gain. Although gain has been introduced by means of electronic amplification in metamaterials\cite{fleury2015invisible} or in piezoelectric materials \cite{willatzen2014acoustic}, this mechanism is difficult to implement for acoustic waves and at low frequencies, for which a more robust approach is required.

In this work we present a mechanism to provide gain and, therefore, to compensate dissipation in mechanical metamaterials based on materials with time-dependent properties. In these materials both the stiffness constant and the mass density are functions of time. 
\rev{
The amplification properties of time dependent media have been of interest  for at least 60 years. The early  studies focused on the parametric amplification in electrical transmission lines  with time varying inductance \cite{Cullen1958,Tien1958} or 
 capacitance \cite{Louisell1958,Honey1960}, and on wave propagation through  dielectric media with time varying properties \cite{Morgenthaler1958,Fante1971}.  
More recent studies have considered  
both time varying mechanical and time varying electromagnetic materials 
\cite{Louisell1958,Lurie2006,hayrapetyan2013propagation,Lurie2016,Milton2017,Mattei2017}. 
Despite the wide interest in these materials, to the best of the authors' knowledge the effects of dissipation on wave amplification in time varying  media have so far not been considered.  
A simplified model for the effect of a resistance element on amplification in transmission line devices is to replace the system by a single degree of freedom RLC circuit with time varying capacitance \cite{Louisell1958}.   This is a useful and instructive model, which is repeated in this work in the context of elasticity, but  it is important to note that it is not directly related to wave amplitudes.  
}
While time dependent media can be thought of as a difficult or nearly impossible to  realize, it has to be taken into account that essentially they are tunable materials which can be quickly reconfigured. The domain of tunable and reconfigurable acoustic and elastic metamaterials is moving fast towards this direction, so that this concept could be doable in the framework of these structures.

We will derive the general properties of a time-dependent dissipative material, showing that the dissipation can be compensated by the amplification of the fields due to the time-dependent properties. It is found that the fields can either blow up or attenuate exponentially with time, but that there is a special regime in which these effects compensate one another and the wave propagates at constant amplitude through the material. The demonstration is based on the analogy between periodically modulated materials in space and time, and it is valid for any periodic function of the constitutive parameters.   \rev{A  single degree of freedom mechanical model is also considered and compared with the fully dynamic continuum model.  
}
\section{Gain in time-dependent media}
\begin{figure}[ht!]
\begin{center}
\includegraphics[width=8cm]{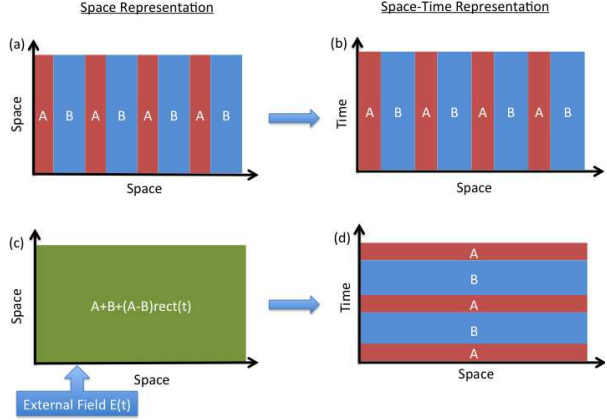}%
\caption{Equivalences between materials with spatial and temporal modulation of their constitutive parameters. Panel (a) shows a classical layered material in space, and panel (b) shows its space-time representation. Panel (c) represents a material whose properties are periodically changed in time by means of an external stimulus, panel (d) is its space-time representation, which shows that it is a rotated version of panel (b).}%
\label{fig:schematics}
\end{center}
\end{figure}

The analogy between spatial and temporal modulation is represented in Figure \ref{fig:schematics}: panel (a) shows a classical layered material with alternating layers of material A and B (a one-dimensional phononic crystal), and panel (b) shows its space-time representation, where we see that the properties (stiffness and mass density) remain constant in time but not in space. Now, let us assume that we have a medium whose material properties are sensitive to some external stimulus such as an electric or magnetic field, an applied stress or even temperature. If this external stimulus $E(t)$ changes with time $t$, the properties of the material will be time-dependent, as shown in panel (c), which represents a material in which the properties change from A to B periodically. Panel (d) shows the space-time representation of this material, where the properties are time-dependent but constant along the space coordinate. The ``space'' representation of these two materials shows two completely different pictures (panels (a) and (c)), however the ``space-time'' representation shows a clear equivalence between these two problems.

The above equivalence between the spatial and temporal modulation of the materials is more evident from the  equation for elastic waves. The one dimensional wave equation for an inhomogeneous elastic material with mass density $\rho(x)$ and stiffness constant $C(x)$ is given by
\dbeq
\frac{\partial }{\partial x}\left(C(x) \frac{\partial u}{\partial x}\right)=\rho(x)\frac{\partial^2u }{\partial t^2},
\deeq
with $u=u(x,t)$ being the $x$ component of the displacement vector. If the properties of the material change in time but not in space, the above wave equation is
\dbeq
\frac{\partial }{\partial t}\left(\rho(t) \frac{\partial u}{\partial t}\right)=C(t)\frac{\partial^2u }{\partial x^2},
\deeq
from which it is clear that there is a direct relationship between the solutions of the equations for space and time modulations.  Therefore if we know the solution for the spatial modulation, we can obtain the solution for the temporal modulation by exchanging the roles of $\rho$ and $C$.

\begin{figure}[ht!]
\begin{center}
\includegraphics[width=8cm]{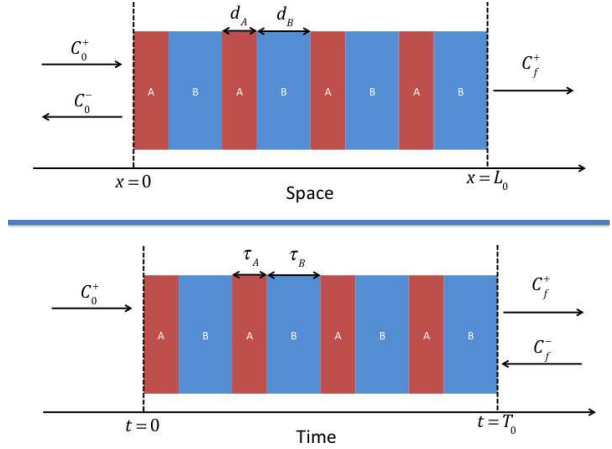}%
\caption{Comparison of the excitation of transmitted and reflected waves by finite layered materials in space (upper panel) and time (lower panel). The position of the reflected wave before (upper panel) or after (lower panel) the slab obeys causality, and it is the responsible of the gain in energy for the temporal layered material (see text for further details).}
\label{fig:ST-reflec}
\end{center}
\end{figure}

Despite the formal analogy between spatial and temporal modulation of materials, there is a fundamental difference between these two situations concerning boundary and initial conditions. This difference manifests itself when we compare the transmission and reflection by a discontinuity in the material in space or time. The process is illustrated in Fig. \ref{fig:ST-reflec}: In the upper panel we see the classical spatial transmission and reflection process in   a layered material, while  the lower panel shows the analogous situation in time.

The upper panel of Fig. \ref{fig:ST-reflec} shows the process of reflection and transmission by a spatial discontinuity: a wave traveling through a given material arrives from the left, encounters  the discontinuity (a layered material in this example), and  a reflected wave is then excited, traveling backwards along the $x$ direction; also, a transmitted wave appears at the other side of the slab, travelling forward in the $x$ direction.

The lower panel of Fig. \ref{fig:ST-reflec} shows the equivalent situation in time: a wave is traveling through a given material in which, due to the application of a periodic temporal external stimulus, from $t=0$ to $t=T_0$ its properties oscillate between two values labelled as $A$ and $B$, to finally rest at its initial state. However the ``position'' of the waves in the schematics is different, since for $t<0$ we have only one wave traveling forward along the $x$ direction; obviously the layered material in time cannot excite a wave traveling ``backwards'' in time. The reflected wave appears after the modulation period, for $t>T_0$, and is in fact a wave traveling backwards in $x$ (it cannot travel backwards in time), so that the result of the modulation in time is the excitation of two waves traveling in opposite directions along the material, as before, but the different position of the reflected wave in the schematics will be the key to understanding the energy gain in the process.

The consequence of this distinction becomes clearer if we use layer theory, which relates the amplitudes of the incoming $C_i^+$ and out-coming $C_i^-$ waves before ($i=0$) and after ($i=f$) the layered structure by means of the scattering matrix $M$, so that, for the spatial case we have
{
\dbeq
\left( \begin{matrix} C_f^+ \\ C_f^-  \end{matrix}\right)=
\left( \begin{matrix} M_{11}^S &  M_{12}^S \\  M_{21}^S & M_{22}^S\end{matrix} \right)
\left( \begin{matrix} C_0^+ \\ C_0^-  \end{matrix}\right).
\deeq }
For an incident wave coming from the left $C_f^-=0$, and the reflection and transmission coefficients are defined as
{
\subeqs{
r_S&\equiv\frac{C_0^-}{C_0^+}= -\frac{M_{21}^S}{M_{22}^S} , \\
t_S&\equiv\frac{C_f^+}{C_0^+}=\frac{1}{M_{22}^S}		
}}
The determinant of the $M$ matrix is unitary, and reciprocity shows that $M_{11}=M_{22}^*$, and $M_{21}=M_{12}^*$, relations that imply $|r_S |^2+ |t_S |^2=1$, that is, there is conservation of energy in the spatial case.

The picture is different for the temporally layered material, where the reflected wave corresponds to the amplitude $C_f^-$. Layer theory is applied likewise, thus
{
\dbeq
\left( \begin{matrix} C_f^+ \\ C_f^-  \end{matrix}\right)=
\left( \begin{matrix} M_{11}^T &  M_{12}^T \\  M_{21}^T & M_{22}^T\end{matrix} \right)
\left( \begin{matrix} C_0^+ \\ C_0^-  \end{matrix}\right)
\deeq}
and {using $C_0^-=0$} the reflection and transmission coefficients are given by
{\subeqs{
r_T&\equiv\frac{C_f^-}{C_0^+}=M_{21}^T , \\
t_T&\equiv\frac{C_f^+}{C_0^+}=M_{11}^T
 } }
The transfer matrix $M^T$ is obtained directly from $M^S$ by changing $C\to\rho$, as discussed before, so that the unitarity and reciprocity relationships will be identical, and it can be easily shown that $|r_T |^2+|t_T |^2\geq 1$ (in fact, $|t_T |^2\geq 1$), that is, there is  increased wave energy. This gain of energy can be understood from the equivalence in the spatial case: since the values for the reflection and transmission coefficients have to be equal or lower than 1, we have that $|M_{22} |=|M_{11} |\geq1$ for both matrices $\bf{M}^T$ and $\bf{M}^S$.

Interestingly, we see that the transmitted energy in the temporal case is the inverse of the transmitted energy in the spatial case. The role of the mass density and the stiffness constant are interchanged between the two situations, which changes the elements of the matrix $\bf{M}$. The most important consequence is that, for a layered material of $N$ periods, when waves propagate at the frequency of the band gap typical of periodic structures, the amplitude of the transmitted wave decreases exponentially with the number of layers, so that its equivalent temporal crystal will have an exponentially increasing gain of energy when the selected wavenumber lies in the band gap. As the number of periods becomes larger, the transmitted energy blows up and the material becomes unstable, unless   the modulation ceases. Therefore, the   stability condition for an infinitely oscillatory medium is that the parameter oscillations are not   strong enough to open a band gap in the dispersion curve.

The above effect can be quantified by means of  layer theory, which shows that the $M$ matrix of a $N$ layer material is given by \cite{bendickson1996analytic}
\dbeq
\bm{M}_N^T=\bm{M}^T\frac{\sin N\Omega\tau}{\sin \Omega\tau}-\bm{I}\frac{\sin (N-1)\Omega\tau}{\sin \Omega\tau}
\deeq
where $\Omega=\Omega(k_0)$ defines the dispersion curve of the infinite periodic material for spatial wavenumber $k_0$. Clearly, within the band gap the element $M_{11}$ has the form $e^{N\tau \Im(\Omega)}$, which grows  exponentially with the number of periods $N$. Therefore, a periodically modulated material will be unstable if its (temporal) band structure presents a band-gap, unless the modulation is of finite duration, in which case  it will act simply as an amplifier.
\section{Loss compensation in time-dependent media}
When dissipation is introduced into the system, the space-time analogy is no-longer valid, and the effect of gain is less evident. The main difference in the spatial case is that dissipation breaks the time reversal symmetry, which means that the material is non-reciprocal in time and the transfer matrix $M^S$ is no longer unitary. Although dissipation is a complex phenomenon with a strong dependence on frequency, the most common assumption in elasticity is to propose a complex stiffness constant directly proportional to the frequency, {$C\to C-i\omega \eta$}, whose origin is the assumption that viscous forces are proportional to the velocity. This  is equivalent to the following time-dependent constitutive equations
\subeqs{
\frac{\partial \sigma}{\partial x}&=\frac{\partial }{\partial t}\left(\rho(t)\frac{\partial u}{\partial t}\right) ,\\
\sigma&=C(t)\frac{\partial u}{\partial x}+\eta(t)\frac{\partial^2u}{\partial x\partial t}
}
where as before $C,\rho$ and the viscosity coefficient $\eta$ are time-dependent. With this model of dissipation, regardless of the temporal dependence of the constitutive parameters, it can be shown that the transfer matrix is given by (see Appendix \ref{app:TD}, equation \eqref{24})
\dbeq
\bm{M}=e^{-\Gamma k_0^2}\widehat{\bm{M}},
\deeq
where the $\Gamma$ factor is given by
\dbeq
\Gamma=\frac{1}{2}\int_{0}^{T_0}\frac{\eta(t)}{\rho(t)}\dd t,
\deeq
and the matrix $\widehat{\bf{M}}$ satisfies unitarity and reciprocity.

The dissipation of the system is described by the exponential factor $e^{-\Gamma k_0^2}$; however this dissipation can be compensated by the elements of the matrix  $\widehat{\bf{M}}$, which is unitary and, therefore,  contributes to the gain of the system. For the specific case of a periodically modulated material, the reflection and transmission coefficients are given by
{\subeqs{\label{eq:rtN}
r_T&=e^{-\Gamma k_0^2}\, \frac{\sin N\Omega \tau}{\sin \Omega \tau} \widehat{M}_{21}^T , \\
t_T&=e^{-\Gamma k_0^2}\left(\frac{\sin N\Omega \tau}{\sin \Omega \tau} \widehat{M}_{11}^T-\frac{\sin (N-1)\Omega \tau}{\sin \Omega \tau}\right)
} }
where now
\dbeq
\Gamma=\frac{N\tau}{2}\left<\frac{\eta(t)}{\rho(t)}\right>
\deeq
with $\langle\cdot\rangle=1/\tau\int_0^\tau \dd t$ being the average in the temporal unit cell $\tau$. The above equations clearly establish the conditions for compensating the  dissipation in the material. If the dispersion curve $\Omega=\Omega(k_0)$ is real, i.e., there is no band gap, all the contributions of the unitary matrix $\widehat{\bf{M}}$ are oscillatory in $N$, and as the number of periods (modulation time) increases the amplitude of both the transmitted and reflected wave decreases because of the exponential factor $e^{-\Gamma k_0^2}$. If the modulation of the parameters is strong enough to open a band gap, the argument in the sinusoidal terms in equations \eqref{eq:rtN} becomes complex and the sine becomes the hyperbolic sine, with an exponentially dominant term as $N$ increases, therefore both the transmission and reflection coefficients have terms of the form
\dbeq
e^{-\Gamma k^2_0}\sin{N\Omega\tau}\approx \exp\left(-{\frac{N\tau}{2}\left<\frac{\eta(t)}{\rho(t)}\right>k_0^2+N\tau\Im(\Omega)}\right)
\deeq
Since both the decaying and growing factors are proportional to $N\tau$,  this exponential term will be compensated and set constant if the condition
\dbeq\label{4733}
\Delta(k_0)\equiv\frac{1}{2}\left<\frac{\eta(t)}{\rho(t)}\right>k_0^2-\Im(\Omega)=0
\deeq
is satisfied, and the transmitted energy will be stable with the modulation time $T_0$, since all the exponential terms (the decaying and the growing ones) have disappeared from the expressions. The energy will therefore propagate along the material without dissipation or amplification. The quantity $\Delta(k^0)$ is therefore the parameter determining the stability of the material. If there is a frequency region where this quantity is negative, the material will be unstable since the energy will blow up exponentially with the number of periods $N$.  This parametric amplification can also be used to gain energy in a controllable way, using the fact that the dissipation $\Delta$ can be a small quantity.    
\rev{ It is interesting to compare the stability condition with the analogous criterion for a single degree of freedom damped oscillator with time varying parameters, see Appendix \ref{2=3}.
}

\begin{figure}[ht!]
\begin{center}
\includegraphics{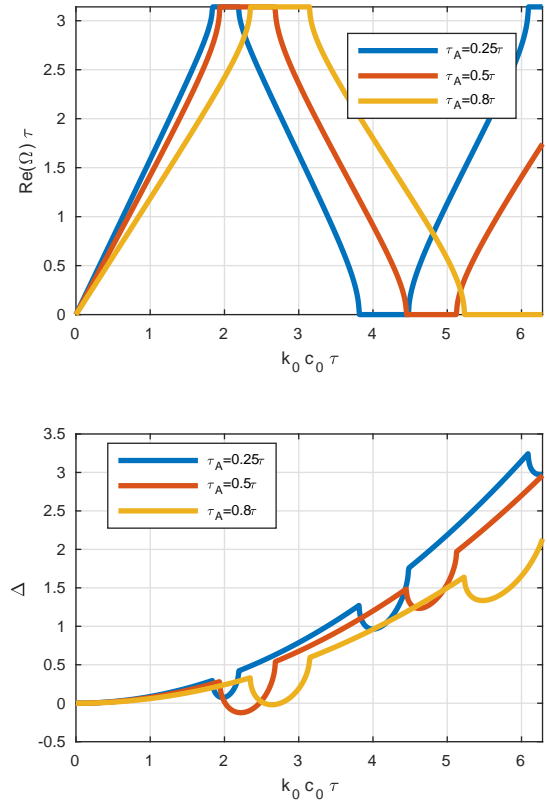}%
\caption{Top: Real part of the dispersion curve $\Omega=\Omega(k_0)$ for a two layer time-dependent medium for different values of $\tau_A$. Bottom: The parameter $\Delta$  of Eq.\ \eqref{4733} as a function of $k_0$ for three different values of $\tau_A$. The regions of $\Delta$ positive, zero or negative correspond to the situations in which the field is dissipated, stabilized or amplified, respectively.}
\label{fig:bands}
\end{center}
\end{figure}

The above results are now illustrated via some numerical examples. Further details regarding the calculations and the expressions employed can be found in the Appendix.

Figure \ref{fig:bands} shows the dispersion curve $\Omega=\Omega(k_0)$ for a two-layer periodic material of time period $\tau= \tau_A +\tau_B$ with elastic properties
$\{\rho_A, C_A, \eta_A\} =\{1,1,0.1\}$
during the time $\tau_A$ and
$\{\rho_B, C_B, \eta_B\} =\{1,3,0.2\}$
in the remaining part of the period  $\tau_B=\tau-\tau_A$. Figure \ref{fig:bands} shows the behaviour for $\tau_A=0.25\tau, 0.5\tau$ and $0.8\tau$. The lower panel shows the parameter $\Delta$ of Eq.\ \eqref{4733} as a function of wavenumber $k_0$.
The curves deviate from parabolic shape only within the band gaps  where $\Im(\Omega)\neq 0$. Observe that for  $\tau_A=0.5\tau$ there is a region for which $\Delta<0$, which   means that the field will be amplified as a function of the number of periods $N$ of the temporal modulation, while for $\tau_A=0.8\tau$ there is a region with $\Delta=0$, so that the field will be stabilized here, despite the fact that the material is dissipative.

\begin{figure}[ht!]
\begin{center}
\includegraphics{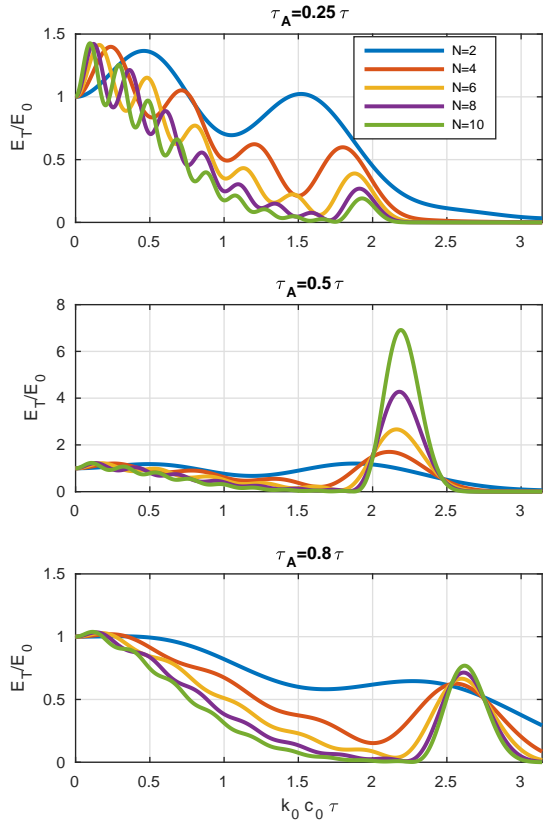}%
\caption{Total energy gain after a periodic modulation of the material for different values of the number of periods $N$. Simulations are shown for $\tau_A=0.25\tau$ (top), $\tau_A=0.5\tau$ (middle) and $\tau_A=0.8\tau$ (bottom), displaying  dissipation, gain and stabilization, respectively.}
\label{fig:energy}
\end{center}
\end{figure}

Figure \ref{fig:energy} shows the gain in energy $E_T/E_0=|r_T|^2+|t_T|^2$ after the modulation of the material's properties for the system of Figure \ref{fig:bands}. Results are shown for different values of the number of periods $N$, and for the different values of $\tau_A$. Clearly, for $\tau_A=0.25\tau$ there is a progressive dissipation of energy as a function of $N$ (upper panel), while for $\tau_A=0.5\tau$ the energy increases as a function of $N$ within the band gap (mid panel). Finally, for $\tau_A=0.8\tau$ there is a situation of stabilization since within the band gap the energy tends to be stable as a function of $N$. It must be pointed out that these three situations depend only on the modulation period $\tau_A/\tau$, which is straightforward to change in practice since it will be the duration for which the external stimulus is in one state or the other, so that the situation of gain-dissipation-stabilization can be externally controlled in these materials.

\begin{figure}[ht!]
\centering
\includegraphics{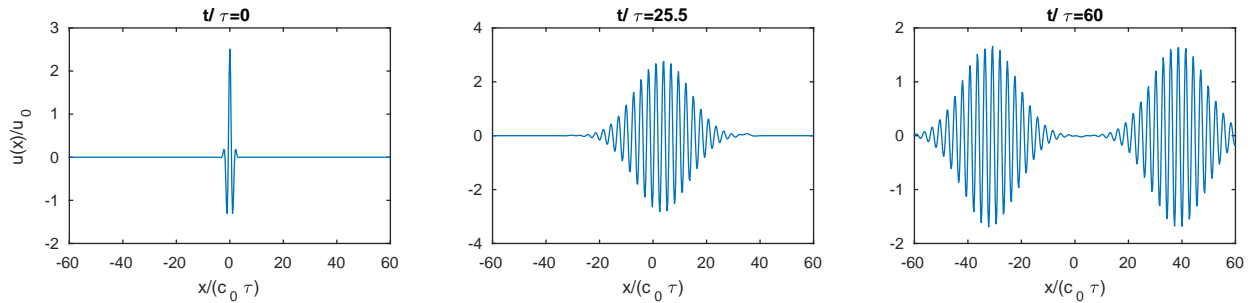}%
\caption{Time evolution of a short pulse in a time-dependent material. It is clear how after the modulation two pulses propagating along opposite directions have been created and amplified.}
\label{fig:TD}
\end{figure}


Figure \ref{fig:TD} shows the time evolution of a Gaussian pulse in a time dependent material under the condition of gain, i.e., $\tau_A=0.5\tau$. The  pulse is chosen to have central frequency at the peak of gain shown in the mid panel of Figure \ref{fig:energy}.
The left, middle and right  panels of Figure \ref{fig:TD} show  the initial pulse,   the response  after the temporal modulation has begun, and the  excitation of the reflected and transmitted wave packets when the modulation ceases, respectively. The full time evolution for this configuration is presented in the supplementary movie \footnote{\url{https://www.dropbox.com/s/2k7r4lwfanh1fub/time_evolution.avi?dl=0}}, which clearly illustrates how  the wavepacket is strongly localized in space during the amplification process, and subsequently propagates after the modulation has stopped. The example shown here corresponds to $N=25$, but the supplementary movie \cite{Note1} illustrates the response for $N=75$, in which the amplification is more evident.
\section{Summary}
In summary, we have presented a general theory for time dependent media showing that  in the absence of dissipation these materials display  gain when the modulation of the parameters is of  finite duration. For continuous and periodic temporal modulation of the material's properties,
 energy  blows up exponentially in  band-gaps, indicating  the possibility  of material instability. By extending the theory  to consider   realistic dissipative materials we have shown that  the energy can decrease or increase exponentially, depending on a  balance equation  which relates the band gap growth to the average value of dissipative parameters. A general equation describes the condition under which the gain due to the band gap and the losses due to dissipation compensate each another, so that the material, despite being dissipative, maintains constant  energy. These results   are valid for any type of time modulation, and  can provide  the basis for the design of loss-compensated metamaterials and devices.

\section*{Acknowledgements}
Work partially supported by the National Science Foundation under award EFRI 1641078, the Office of Naval Research under MURI Grant No. N00014-13-1-0631 and Grant No. N00014-17-1-2445, the LabEx AMADEus (ANR-10- 444 LABX-42) in the framework of IdEx Bordeaux (ANR-10- 445IDEX-03-02), and the Engineering and Physical Sciences Research Council, UK, under grant EP/L018039/1.

\appendix

\section{Time dependent medium}
\label{app:TD}
\subsection{Basic equations with space time parameters}

The field variables are displacement $u(x,t)$, velocity $v=\partial_t u$ and strain 
$\varepsilon =\partial_x u$.   
The equilibrium equation and the stress constitutive relation are 
\bse{1}
\bal{1a}
 \partial_x\sigma &=\partial_t(\rho v), 
\\
\sigma &=\C \varepsilon   + \eta \partial_x v  
\label{2}
\eal
\ese
where the material parameters are density $\rho (x,t)$, stiffness $\C(x,t)$ and viscosity $\eta (x,t)$. 
Equations \eqref{1}   together give the pointwise energy balance
\beq{3}
\partial_t \big( \frac 12 \rho v^2 + \frac 12 \C \varepsilon^2\big)  + \partial_x (-v\sigma)   
= \frac 12 \varepsilon^2 \partial_t \C -\frac 12 v^2 \partial_t \rho - \eta (\partial_x v)^2   .
\eeq
The right hand members in \eqref{3} are clearly producers of energy, the first two could be either sources or drains, while the final term is the expected viscous loss.

\subsection{Space harmonic solution}
We focus on time dependent material properties: $\rho= \rho (t)$,  $\C=\C(t)$ and  $\eta = \eta (t)$. 
Assume that the variables have separate space and time dependence
\beq{4}
u(x,t) = \Re \, u(t) e^{ik_0x}, 
\eeq
where $u(t)$ is a complex-valued quantity  and $k_0$ is real-valued and positive.  Similar expressions follow for the other variables, and from here on we consider $v(t)= \partial_t u(t)$, $\varepsilon(t) = ik_0 u(t)$ and $\sigma(t)= ik_0 \big( \C u(t) + \eta v(t)\big)$ as complex quantities with the space harmonic factor $e^{ik_0x}$ omitted but understood, analogous to how we  consider time harmonic motion.  

Equations  \eqref{1}  become
\beq{8}
\partial_t {\bf U}
= -ik_0 
\begin{bmatrix}
0 & \rho^{-1}
\\
\C  & -ik_0 \eta\rho^{-1}
\end{bmatrix}{\bf U}
\eeq
where 
\beq{-6}
{\bf U}(t) \equiv 
   \begin{pmatrix}
\varepsilon \\ - \p
\end{pmatrix} \ \ \text{and } \
\p (t) = \rho (t) v  (t)
\eeq
is the momentum.  The propagator, or transfer matrix, for ${\bf U}$ is not unitary.  The connection with unitarity and reciprocity can be made by first defining the speed, impedance and non-dimensional viscosity, 
\beq{10-0}
c= \sqrt{{\C}/{\rho}}, \ \ 
z= \rho c, \ \ 
\zeta = \frac{k_0\eta}{2z}  .
\eeq
Consider  wave solutions of Eq.\  \eqref{1} for constant material properties $\rho$, $C$ and $\eta$ of the form $(u,\sigma) = (u_0,,\sigma_0) \exp\{ik_0(x-\lambda_0 ct)\}$. The non-dimensional frequency  $\lambda_0$ satisfies
\beq{2-55}
\lambda_0^2 + 2i\zeta \lambda_0 -1 = 0, 
\eeq
which implies that propagating waves in $t>0$ occur only for  $\zeta <1$; otherwise the wave is exponentially decaying with time.   Hereafter is it assumed that \rev{the damping factor} $\zeta$ is less than the critical value of unity. 

Equation \eqref{8}  can now be rewritten 
\beq{9}
\partial_t {\bf V}(t)
= {\bf Q} (t){\bf V}(t)
\eeq
where 
\beq{10}
\begin{aligned}
{\bf V}(t)& = e^{\int^t_0 k_0 c\zeta \dd t} \, {\bf U}(t)  ,
\\
{\bf Q}&=  - ik_0 c\sqrt{1-\zeta^2} {\bf A}, 
\ \
{\bf A} &= \frac 1{\sqrt{1-\zeta^2}} 
\begin{bmatrix}
i\zeta  & z^{-1}
\\
z & -i\zeta
\end{bmatrix}   .
\end{aligned}
\eeq
The property  ${\bf A}^2 = {\bf I}$ where ${\bf I}$ is the 2$\times$2 identity matrix leads to the usual unitary properties for the propagator and other related results, below.   The restriction $\zeta <1$ implies that the effects of dissipation are described entirely through the exponential term in Eq.\ \eqref{10}. 

We next consider the propagator and   transfer  matrices using well known methods \cite{Pease}. 
The following results are for arbitrary time dependent ${\bf Q}(t)$ of the form defined in \eqref{10}. 

\subsection{Propagator and transfer matrices}

Let ${\bf P}(t)$ be the 2$\times$2 matrix solution of the differential initial value problem
\beq{211}
\partial_t {\bf P}(t)
= {\bf Q} (t){\bf P}(t), 
\ \ {\bf P}(0) = {\bf I}. 
\eeq
The propagator matrix $\bf P$ relates the state vector ${\bf V}(t) $ at one time with that at another, 
$
{\bf V}(t) = {\bf P}(t)
{\bf V}(0) $, and it has the usual   properties of an undamped propagator, such as determinant of one. 
The actual "damped" propagator for ${\bf U}(t)$ follows from \eqref{10} as $e^{- k_0^2 \Gamma (t)} {\bf P}$, since 
\beq{1-}
{\bf U}(t) 
= e^{- k_0^2 \Gamma (t)} {\bf P}(t)
{\bf U}(0)  
\eeq
and 
\beq{00}
\Gamma (t) = \frac 12 \int_0^t \frac{\eta(t)}{\rho(t)} \dd t. 
\eeq
For future reference define
\beq{21}
{\bf J} = 
\begin{bmatrix}
0  &1
\\
1 & 0
\end{bmatrix},  
\ \ 
{\bf Z} (t)= 
\begin{bmatrix}
1  &1
\\
z & -z
\end{bmatrix}.
\eeq 

We assume  a finite layer of thickness (time) $T$ with uniform properties  outside.  Then 
	\beq{9-}
{\bf W}(T) 
={\bf M}  
  {\bf W}(0) 
\ \ \text{where } \ {\bf W}(t) \equiv   \begin{pmatrix}
C^+ (t) 
\\
C^- (t) 
\end{pmatrix} .
\eeq
  A forward (backward) traveling wave satisfies 
$-\p = z \varepsilon$ ($-\p = -z \varepsilon$).  The forward and backward components  ${\bf W}$ are connected to the state vector ${\bf U}$ via the impedance matrix by ${\bf U}={\bf Z}{\bf W}$ so that ${\bf M}= {\bf Z}^{-1}(T) {\bf P}(t) {\bf Z}(0)$. The periodicity assumption implies 
${\bf Z}(T)= {\bf Z}(0)$ so that 
\beq{24}
{\bf M} =  e^{-\Gamma k_0^2} \, \widehat{\bf M} 
\ \ \text{where }\ 
 \widehat{\bf M} =
{\bf Z}^{-1}(0) {\bf P}(T) {\bf Z}(0) 
\eeq
where the damping constant is  
\beq{-34}\Gamma \equiv \Gamma (T). 
\eeq
Note: the term $e^{-\Gamma k_0^2} $ comes from the definition \eqref{9-}.  
 
The following identities are readily derived: 
\bse{22}
\bal{222}
{\bf Q}^\dagger &= -{\bf J}{\bf Q}{\bf J}, 
\\
{\bf P}^{-1}(t) & =  {\bf P}(-t)= {\bf J}{\bf P}^\dagger(t){\bf J},
\\
\det {\bf P}(t) &=\det \widehat{\bf M} = 1,
\eal
\ese
where $^\dagger$ is the Hermitian transpose (transpose plus conjugate).
These imply  properties for the  elements of $\bf P$ and $\bf M$ (and similar ones  for $\widehat{\bf M} $ as for $\bf M$):
\beq{23}
\begin {aligned}
P_{11}^*&=P_{11}, \ P_{22}^*=P_{22}, \ P_{12}^*=-P_{12}, \ P_{21}^*=-P_{21}, \\
M_{11}^*&=M_{22}, \ M_{12}^*=M_{21}, \\
\text{where } \
{\bf P}(t) &= \begin{bmatrix}
P_{11}&  P_{12}
\\
P_{21}&  P_{22}
\end{bmatrix} ,
\ \ 
{\bf M} = \begin{bmatrix}
M_{11}&  M_{12}
\\
M_{21}&  M_{22}
\end{bmatrix} 
\end {aligned}
\eeq
and  $^*$ indicates complex conjugate.

According to the definition \eqref{9-}, the reflection and transmission coefficients for the time-vary medium can be found using $C^-(0)=0$ as 
\bse{33}
\bal{331}
r_T &= \frac{C^-(T)}{C^+(0)} = M_{21},
\\
t_T &= \frac{C^+(T)}{C^+(0)} = M_{11} .
\eal
\ese
These  satisfy 
\beq{332}
|t_T|^2 - |r_T|^2 = e^{-2\Gamma k_0^2}  
\eeq
  implying  magnification of the forward traveling wave ($|t_T| \ge 1$) in the absence of damping $(\zeta = 0)$.  

\subsection{Piecewise constant regions}
 For instance, if ${\bf Q}$ is constant as a function of $t$, then the solution of \eqref{211}
\beq{11}
{\bf P}(t)  = 
e^{ {\bf Q} t} =
\cos \phi\, {\bf I}  - i \sin \phi\,  {\bf A}
\ \ \text{with } \ 
\phi  =  k_0 c\sqrt{1-\zeta^2} \,T .
\eeq
For a layer $\{c,z,\zeta \}$ sandwiched in time by the uniform medium with impedance $z_0$, 
\beq{2=4}
{\bf M} =  e^{-\Gamma k_0^2} \,
\bigg(
\cos\phi \, {\bf I} - \frac{i\sin\phi}{2\sqrt{1-\zeta^2}}
\begin{bmatrix}
\frac z{z_0} +\frac{z_0} z & \frac z{z_0} -\frac{z_0} z +i2\zeta
\\
 \frac  {z_0}z -\frac z{z_0}  +i2\zeta & - \big(\frac z{z_0} +\frac{z_0} z  \big)
\end{bmatrix} .
\bigg)
\eeq
The reflection and transmission amplitudes follow from eqs.\ \eqref{33} and \eqref{2=4} as
\bse{3=3}
\bal{2323}
|r_T|^2 &=   e^{-2\Gamma k_0^2} \, \frac{\sin^2\phi}{1-\zeta^2} \Big[ \frac 14 \Big( \frac z{z_0} -\frac{z_0} z\Big)^2 +\zeta^2 
  \Big]  ,
\\
|t_T|^2 &=  e^{-2\Gamma k_0^2} \, \Big\{ \frac{\sin^2\phi}{1-\zeta^2} \Big[ 
 \frac 14 \Big( \frac z{z_0} -\frac{z_0} z\Big)^2 +\zeta^2\Big] +1\Big\} .
\eal
\ese

\subsection{Bloch-Floquet}
If the time-modulation is $T-$periodic  then there exist Bloch-Floquet solutions of the form $u(t+T) = u(t)e^{- i\omega T}$, with  similar expressions for the other variables.  The frequency $\omega$ satisfies the eigenvalue condition 
\beq{08}
\det \big(e^{- k_0^2 \Gamma} {\bf P}(T) - e^{-i\omega T}{\bf I}\big)=0 .
\eeq 
Therefore
\beq{232}
\omega = \pm \Omega -i \frac{ k_0^2 \Gamma}{T}
\eeq
where $\Omega$ is the undamped Bloch-Floquet frequency, 
\beq{0-1}
\cos \Omega T =  \frac 12   \, \tr{\bf P}(T) .
\eeq
Note, the frequency $\omega$ is complex valued when  damping is present. 
 In the absence of damping ($\eta = 0$) the frequency is real-valued in the pass bands and complex in stop bands. 

\subsection{Example: two-layer system}
The materials are 1 and 2, of duration $t_1$ and $t_2$, $t_1+t_2 =T$, then \eqref{0-1} is 
\beq{0-2}
\cos \Omega T = \cos k_0 c_1 t_1 \cos k_0 c_2 t_2
-\frac 12 \Big(\frac {z_1}{z_2} +\frac{z_2}{z_1}\Big) \sin k_0 c_1 t_1 \sin k_0 c_2 t_2  .
\eeq
This can be used to look at the response within band-gaps. 

For instance, suppose the speeds and temporal thicknesses are the same in both layers $(c_1=c_2=c, \, t_1=t_2=T/2)$.  The first band gap, defined by 
$ \cos \Omega T <-1$, is 
\beq{121}
k_0 \in \frac 1{cT} \, \Big( \pi - \theta,  \pi + \theta\Big)
\ \ \text{where }\ \cos\frac{\theta}2 = \frac 2 { \sqrt{\frac{z_1}{z_2}} +\sqrt{\frac{z_2}{z_1}} }, \ \ 0<\theta <\pi .
\eeq
The frequency $\Omega$ has its largest imaginary value in the middle of the band gap, $k_0cT = \pi$, where 
\beq{34-}
 \Omega  = \frac{\pi}T + \frac iT\, \Big| \log \frac{z_1}{z_2}\Big|  .
\eeq
Thus, the loss from damping compensates the gain of the  temporal modulation if the two are  matched according to  $\Delta(k_0) = 0$, which in this case becomes
\beq{50234}
\frac{1}{2}\left<\frac{\eta(t)}{\rho(t)}\right>k_0^2
= \frac 1T\, \Big| \log \frac{z_1}{z_2}\Big| .
\eeq

\subsection{Energy evolution} 

Averaging eq.\ \eqref{3} over a wavelength $2\pi/k_0$, yields a purely time-dependent energy result 
\beq{12}
\partial_t \Big(  \frac 12 \C |\varepsilon|^2 +\frac 12 \rho^{-1} |\p|^2 \Big)  
= \frac 12 |\varepsilon|^2 \partial_t \C +\frac 12 |\p|^2 \partial_t \rho^{-1} - \eta \frac{k_0^2}{\rho^2} |\p|^2 . 
\eeq
This form of the energy balance is  instructive since it involves the quantities $\varepsilon$ and $\p$ that vary smoothly in time according to equation \eqref{8}.  It is also interesting to note that 
$\varepsilon$, $\p$ are the dual variables to $v$, $\sigma$ which define the state vector in the spatially modulated material.  The two sets of variables appear on an equal footing in the Willis equations \cite{Norris12}.

 Equation \eqref{12}  indicates  that the energy is not necessarily smoothly varying, since instantaneous jumps in $C$ and/or $\rho$ lead to instantaneous non-zero changes in the total energy.    In order to understand this further, for the moment we ignore damping and consider a uniform medium with properties $\{C_0, \rho_0\}$ that instantaneously switches to  $\{C_1, \rho_1\}$ at time $t=0$.   Equation \eqref{12} with $\eta = 0$ implies the energy equation 
\beq{12_1}
 E = E_0 + \frac 12\Big(  |\varepsilon_0|^2 (C_1-C_0)+  |\p_0|^2 (\rho_1^{-1}- \rho_0^{-1})
\Big) \, H(t) 
\eeq
where $E = \frac 12 \C |\varepsilon|^2 +\frac 12 \rho^{-1} |\p|^2 $ is the total energy, 
$E_0$, $\varepsilon_0$ and $\pi_0$ are the values just before  $t=0$,  
 and $H(t)$ is the Heaviside step function. According to  equation \eqref{12_1} the instantaneous change could  increase or decrease the total energy.  For instance, if the wave before the switch is propagating in one direction,  $\pi_0 = \pm z_0 \varepsilon_0$, then the energy increases if 
$ \frac{z_1}{z_0}+ \frac{z_0}{z_1} > 2     \frac{c_0}{c_1}$, otherwise it decreases, 
where $C_1 = c_1 z_1$, $\rho_1^{-1} = c_1/z_1$

Now suppose that the medium properties revert at time $t=T>0$ to those before the switch at $t=0$, then the subsequent energy for $t>T$ is 
\beq{12_2}
 E_T = E_0 + \frac 12  (|\varepsilon_0|^2-|\varepsilon_1|^2 ) (C_1-C_0)+  \frac 12 (|\p_0|^2 -|\p_1|^2) (\rho_1^{-1}- \rho_0^{-1})
\eeq
where the values 
$\varepsilon_1$ and $\pi_1$    at  $t=T$ are related to those at $t=0$ by the propagator, 
\beq{12_3}
 \begin{pmatrix}
\varepsilon_1 \\ - \p_1
\end{pmatrix}
=  
\begin{bmatrix}
\cos (k_0c_1T)  &  -i z_1^{-1} \sin (k_0c_1T)  
\\
-i z_1 \sin (k_0c_1T)   & \cos (k_0c_1T) 
\end{bmatrix}
 \begin{pmatrix}
\varepsilon_0 \\ - \p_0
\end{pmatrix} .
\eeq
  Hence, 
\beq{12_35}
 E_T = E_0 + \frac {C_0}2  \Big( \frac{z_1^2}{z_0^2}- 1\Big)
\Big( |\varepsilon_0|^2 -  \frac{|\p_0|^2 }{z_1^2}\Big)
\sin^2 (k_0c_1T)
\eeq
implying that the energy can increase or decrease relative to $E_0$ depending on the wave dynamics at $t=0$. 
However, assuming the dynamic state just before the first switch at $t=0$ is a wave travelling in one direction only, i.e. $\pi_0 = \pm z_0 \varepsilon_0$, then the energy after reversion is 
\beq{12_4}
 E_T = E_0 + \frac {E_0}{2}  \Big( \frac{z_1}{z_0}- \frac{z_0}{z_1}\Big)^2 
\sin^2 (k_0c_1T)
  .
\eeq
This is always greater than or equal to the initial energy.  
 
In summary, for wave incidence in one direction on the temporal slab of width $T$:   (i) the evolved energy is  non-decreasing, (ii) it remains constant for any temporal slab width $T>0$ if and only if the impedance remains constant, and (ii) for a given impedance mismatch, the energy increase is maximum for  $T $ such that $\sin^2 (k_0c_1T) = 1$. 

\section{A single degree of freedom model}  
\label{2=3}
\rev{Consider a mass-spring-damper  system with time varying stiffness 
\beq{2=31}
m \ddot u (t) + c\dot u(t) +(K+\Delta K \cos 2 \omega_n )u(t) = 0 
\eeq
where $\omega_n = \sqrt{K/m}$ is the undamped natural frequency. The displacement solution has the form $u(t) = v(t) e^{(\mu - \zeta)\omega_n t}$ where $v(t)$ is periodic and $\zeta = \frac{c}{2\omega_n m}$ is the damping factor.  For small values of $\frac{\Delta K}{K}$ the parameter $\mu$ follows from \cite[\S4.23, Eqs.\ (3) and (4)]{McLachlan}.  In particular 
\beq{2=32}
\Re \mu  \approx \frac{\Delta K}{8K} . 
\eeq
Note that this differs by a factor of two from the analogous result in \cite{Louisell1958}. 
Hence, the solution will grow exponentially if  
\beq{2=33}
 \frac{\Delta K}{8K} > \zeta . 
\eeq
}

\rev{
In order to compare this  with the full wave result of equation \eqref{4733} we note that the 
latter implies exponential growth if 
$\Im(\Omega) >\frac 12 \left<\frac{\eta(t)}{\rho(t)}\right>k_0^2
$.  Assuming time independent wave speed $c = \sqrt{C/\rho}$ and non-dimensional viscosity $\zeta$ of equation  \eqref{10-0}, the growth condition  can be expressed 
\beq{5-33}
\frac{\Im(\Omega)}{\Omega_0} >  \zeta 
\eeq
where $\Omega_0 \equiv ck_0$. In both the simple model \eqref{2=33} and the full wave solution
\eqref{5-33}, the condition for exponential growth involves the non-dimensional damping factor $\zeta$. In the single degree of freedom model it competes with the relative change in stiffness.   However, in the full wave case, the damping competes with an indirect quantity which depends upon the Bloch-Floquet response.   
}


\end{document}